\documentclass[aps, prl, twocolumn, superscriptaddress, amsmath,  tightenlines, longbibliography]{revtex4-1}

\usepackage{dcolumn}
\usepackage{graphicx}
\usepackage{mathrsfs}
\usepackage{subfigure}
\usepackage{booktabs}
\usepackage{amsmath}
\usepackage{physics}
\usepackage{dsfont}
\usepackage{amstext}
\usepackage{amssymb}
\usepackage{amsbsy}
\usepackage{bbm}
\usepackage{amsthm}
\usepackage{graphicx}
\usepackage{color}
\usepackage[colorlinks,citecolor=blue]{hyperref}

\usepackage{url}
\usepackage[colorlinks]{hyperref}
\hypersetup{%
	plainpages=true,
	breaklinks=true,       
	hypertexnames=false,  
	pageanchor=true,
	colorlinks=true,
	linkcolor={blue},
	citecolor={red},
	urlcolor={blue},
	anchorcolor={black}
}

 \makeatletter

\newcommand{\Rmnum}[1]{\expandafter\@slowromancap\romannumeral #1@}
\makeatother

\begin{document}
\title{Higher-Order Weyl-Exceptional-Ring Semimetals}
\author{Tao Liu}
\thanks{These two authors contributed equally}
\affiliation{School of Physics and Optoelectronics, South China University of Technology,  Guangzhou 510640, China}
\author{James Jun He}
\thanks{These two authors contributed equally}
\affiliation{RIKEN Center for Emergent Matter Science, Wako, Saitama 351-0198, Japan}
\author{Zhongmin Yang}
\email[E-mail: ]{yangzm@scut.edu.cn}
\affiliation{School of Physics and Optoelectronics, South China University of Technology, Guangzhou 510640, China}
\affiliation{South China Normal University, Guangzhou 510006, China}
\affiliation{State Key Laboratory of Luminescent Materials and Devices and Institute of Optical Communication Materials, South China University of Technology, Guangzhou 510640, China}
\author{Franco Nori}
\email[E-mail: ]{fnori@riken.jp}
\affiliation{Theoretical Quantum Physics Laboratory, RIKEN Cluster for Pioneering Research, Wako-shi, Saitama 351-0198, Japan}
\affiliation{RIKEN Center for Quantum Computing (RQC), Wako-shi, Saitama 351-0198, Japan}
\affiliation{Department of Physics, University of Michigan, Ann Arbor, Michigan 48109-1040, USA}

\date{{\small \today}}


\begin{abstract}
For first-order topological semimetals, non-Hermitian perturbations can drive the Weyl nodes into Weyl exceptional rings having multiple topological structures and no Hermitian counterparts. Recently,  it was discovered that higher-order Weyl semimetals, as a novel class of higher-order topological phases, can uniquely exhibit coexisting surface and hinge Fermi arcs.  However, non-Hermitian higher-order topological semimetals have  not yet been explored. Here, we identify a new type of topological semimetals, i.e, a higher-order topological semimetal with Weyl exceptional rings. In such a semimetal, these rings are characterized by both a spectral winding number and a Chern number. Moreover, the higher-order Weyl-exceptional-ring semimetal supports both surface and hinge Fermi-arc states, which are bounded by the projection of the Weyl exceptional rings onto the surface and hinge, respectively. Noticeably, the dissipative terms can cause the coupling of two exceptional rings with opposite topological charges, so as to induce topological phase transitions. Our studies open new avenues for exploring novel higher-order topological semimetals in non-Hermitian systems.
\end{abstract}

\maketitle

\textit{Introduction}.---There is growing interest in exploring higher-order topological insulators \cite{PhysRevLett.110.046404, Benalcazar61, PhysRevB.96.245115, PhysRevLett.119.246401,PhysRevLett.119.246402,  PhysRevB.97.241405, Peterson2018, Serra-Garcia2018,   arXiv:1801.10053,  PhysRevLett.120.026801,  TitusSciAdv2018,Zhang2019, Ni2018, Xue2018, arXiv:1801.10050, arXiv:1802.02585,   PhysRevLett.123.216803,  Mittal2019, ElHassan2019, Li2019,    PhysRevResearch.2.033029, PhysRevLett.124.036803, PhysRevB.101.241104,PhysRevLett.124.036803, PhysRevLett.124.063901} and superconductors \cite{PRBXYZhu2018, arXiv:1803.08545,   arXiv:1806.07002, PhysRevLett.121.196801, PhysRevLett.123.177001, PhysRevB.100.205406, PhysRevLett.122.236401, PhysRevB.99.020508, PhysRevLett.122.126402, PhysRevB.100.075415, PhysRevLett.122.187001, PhysRevLett.123.156801, PhysRevB.99.125149, PhysRevX.10.041014,  PhysRevLett.124.227001, PhysRevResearch.2.012060, arXiv:2007.10326}. As a new family of topological phases of matter, higher-order topological insulators and superconductors show an unconventional bulk-boundary correspondence, where a $d$-dimensional $n$th-order ($n \geq 2$) topological system hosts topologically protected  gapless states on its $(d - n)$-dimensional boundaries, such as the corners or hinges of a crystal.  Very recently, the concept of higher-order topological insulators has been extended to 3D gapless systems, giving rise to distinct types of topological semimetal phases with protected nodal degeneracies in bulk bands and hinge Fermi-arc states in their boundaries. Examples include higher-order Dirac semimetals \cite{PhysRevB.98.241103, Wieder2020}, higher-order Weyl semimetals \cite{PhysRevResearch.1.032048, PhysRevLett.125.146401, PhysRevLett.125.266804, Luo_2021,Wei2021}, and higher-order nodal-line semimetals \cite{PhysRevB.99.041301, PhysRevLett.123.186401, PhysRevLett.125.126403}. 

Weyl semimetals exhibit two-fold degenerate nodal points in momentum space, called Weyl points (or Weyl nodes). The Weyl points are quantized monopoles of the Berry flux, and are characterized by a quantized Chern number on a surface enclosing the point \cite{RevModPhys.90.015001}. The nontrivial topological nature of  first-order Weyl semimetals guarantees the existence of surface Fermi-arc states, connecting the projections of each pair of Weyl points onto the surface.   In contrast to first-order Weyl semimemtals, higher-order Weyl semimetals have bulk Weyl points attached strikingly to both surface and hinge Fermi arcs \cite{PhysRevLett.125.146401, PhysRevLett.125.266804}.  

Recently, considerable efforts have been devoted to explore  topological phases in non-Hermitian extensions of topological insulators \cite{PhysRevLett.116.133903,  PhysRevLett.118.040401, PhysRevLett.120.146402, Harari2018,   Xiong2018, arXiv:1802.07964, arXiv:1806.06566,  arXiv:1805.06492,  ShunyuYao2018, YaoarXiv:1804.04672,  PhysRevB.100.054105, Kawabata2019,    PhysRevB.99.235112, PhysRevX.9.041015, PhysRevB.100.144106, PhysRevB.100.035102,  arXiv:1905.02211,    PhysRevLett.123.097701,  PhysRevLett.123.066404,  PhysRevLett.122.237601,  PhysRevB.100.184314,    PhysRevB.100.045141,  PhysRevLett.123.206404, PhysRevB.100.155117, PhysRevLett.123.206404,  Zhao2019, PhysRevB.99.201103, PhysRevB.101.121116, PhysRevLett.124.086801, PhysRevB.101.020201,   PhysRevLett.124.056802, PhysRevB.102.035153, RevModPhys.93.015005,  PhysRevB.101.235150,PhysRevB.102.085151, Xiao2020,Helbig2020,arXiv:2006.01837,arXiv:2102.02230} and semimetals \cite{arXiv:1812.02011, Zhou2018,PhysRevB.99.201107, PhysRevLett.118.045701,PhysRevB.97.075128,PhysRevB.99.075130,PhysRevB.99.041406,PhysRevLett.123.066405,PhysRevB.99.121101, PhysRevB.100.245116,Cerjan2019, PhysRevB.99.041116, PhysRevA.102.023308, PhysRevLett.124.186402,Lee2020,arXiv:2102.05059}, including non-Hermitian higher-order topological insulators \cite{PhysRevLett.122.076801, PhysRevB.99.081302, PhysRevLett.123.073601, PhysRevLett.123.016805,PhysRevB.99.201411, PhysRevB.103.L041102, PhysRevB.102.094305,PhysRevB.103.L041115,arXiv:2104.11260}. Non-Hermiticity originated  from dissipation in open classical and quantum systems    \cite{RevModPhys.93.015005,arXiv:2006.01837}, and the inclusion of non-Hermitian features in topological systems can give rise to unusual topological properties and novel topological phases without Hermitian counterparts. One striking feature is the existence of non-Hermitian degeneracies, known as exceptional points, at which two eigenstates coalesce \cite{NMSahen2019,Gao_2015,PhysRevA.100.062131,arXiv:2006.01837,PhysRevA.104.012205}. The non-Hermiticity can alter the nodal structures, where the exceptional points form new types of topological semimetals   \cite{ PhysRevLett.118.045701,PhysRevB.97.075128,PhysRevB.99.075130,PhysRevB.99.041406,PhysRevLett.123.066405}. Remarkably, a non-Hermitian perturbation can transform a Weyl point into a ring of exceptional points, i.e., Weyl exceptional ring  \cite{PhysRevLett.118.045701}.  This Weyl exceptional ring carries a quantized Berry charge, characterized by a Chern number defined on a closed surface encompassing the ring, with the existence of the surface states. In addition, such a ring is also characterized by a quantized Berry phase defined on a loop encircling the ring.  Weyl exceptional rings show multiple topological structures having no Hermitian analogs in Weyl semimetals \cite{PhysRevLett.123.066405}. Although non-Hermitian first-order  topological semimetals have been systematically  explored, far less is known on non-Hermitian higher-order topological semimetals with hinge states. This leads to a natural question of whether a non-Hermitian perturbation can transform a Weyl point into a Weyl exceptional ring in a higher-order topological semimetal. 

In this work, we investigate non-Hermitian higher-order topological semimetals, where a non-Hermitian perturbation transforms the higher-order Weyl points into Weyl exceptional rings formed by a set of exceptional points. The topological stability of such a ring is characterized by a non-zero spectral winding number. In addition, the Weyl exceptional ring has a quantized non-zero Chern number when a closed surface encloses the ring. This leads to the emergence of  surface Fermi-arc states. Moreover, as a new type of topological semimetals, the higher-order Weyl-exceptional-ring semimetals (HOWERSs) have the hinge Fermi arc connected by the projection of the Weyl exceptional rings onto the hinges. Meanwhile, the dissipative terms can induce topological phase transitions between different semimetal phases. By developing an effective boundary theory, we provide an intuitive understanding of the existence of  hinge Fermi-arc states: the surface states of the first-order topological semimetal with Weyl exceptional rings are gapped out by an additional anticommutative term in a finite wavevector-$k_z$ region. This introduces Dirac mass terms, which have opposite signs between the neighboring surfaces. Therefore, hinge states appear only in a finite $k_z$ region, resulting in Fermi-arc states.

\textit{Hamiltonian}.---We start with the following minimal non-Hermitian Hamiltonian on a cubic lattice
\begin{align}\label{H1}
\mathcal{H}(\mathbf{k}) = ~&\left(m_0 - \cos k_x  - \cos k_y + m_1 \cos k_z \right) s_z \sigma_z  \nonumber \\
& + \left(v_z \sin k_z  + i \gamma\right) s_z +  \sin k_x   s_x \sigma_z  + \sin k_y s_y \sigma_z  \nonumber \\ 
& + \Delta_0  \left(\cos k_x - \cos k_y \right) \sigma_x,
\end{align}
where $\sigma_i$ and $s_i$ are Pauli matrices, and $\gamma$ denotes the decay strength. The Hamiltonian $\mathcal{H}(\mathbf{k})$ can in principle be experimentally realized using dissipative ultracold atoms and topoelectric circuits (see Supplementary Material \cite{SMHighOrderNodalRings}).

The Hamiltonian $\mathcal{H}(\mathbf{k})$ preservers: (1)  time-reversal symmetry $\mathcal{T} = i s_x \sigma_x \mathcal{K}$, with $\mathcal{K}$ being the complex conjugation operator, and (2) the combined charge conjugation and parity  ($\mathcal{CP}$) symmetry $\mathcal{CP} \mathcal{H}^T(\mathbf{k})\left(\mathcal{CP}\right) ^{-1}=-\mathcal{H}(\mathbf{k})$, with $\mathcal{CP} = s_y \sigma_z $. For $\gamma = 0$, the system is a hybrid-order Weyl semimetal, which supports both first-order and second-order Weyl nodes with coexisting surface and hinge Fermi arcs \cite{SMHighOrderNodalRings}.

\begin{figure}[!tb]
	\centering
	\includegraphics[width=8.4cm]{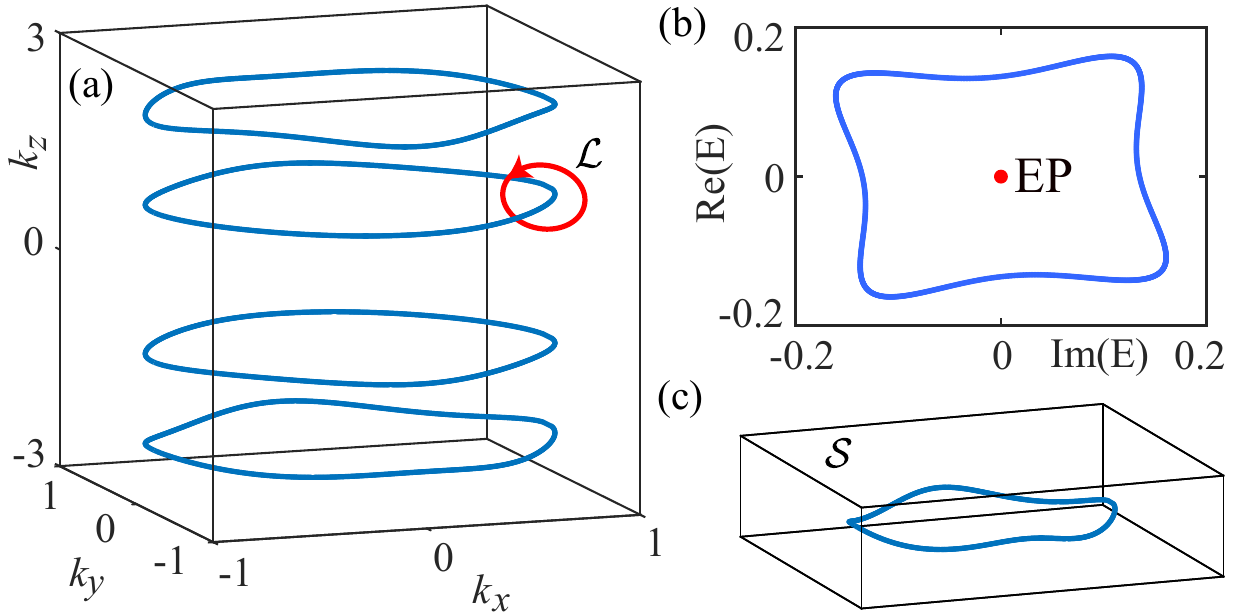}
	\caption{(a) Weyl exceptional rings (blue curves) obtained by solving Eqs.~(S7,S8) in Ref.~\cite{SMHighOrderNodalRings}). They are formed by a two-band coalescence at the energy $E = 0$. The bulk bands of the non-Hermitian Hamiltonian $\mathcal{H}(\mathbf{k})$ in Eq.~(\ref{H1}) exhibit four Weyl exceptional rings along the $k_z$ direction. The parameters used here are: $m_0 = 1.5$, $m_1  = -1$, $v_z=0.8$, $\gamma=0.8$, and $\Delta_0 =0.8$.  The red loop denotes a closed path $\mathcal{L}$  encircling a Weyl exceptional ring. (b) Real parts vs. imaginary parts of eigenenergies along the path  $\mathcal{L}$. The red dots denote exceptional points (EPs).   (c) A Weyl exceptional ring is enclosed by a closed surface of a cuboid.}\label{ERings}
\end{figure}

\textit{Weyl exceptional ring}.---In the presence of the non-Hermitian term $\gamma$, two of the bulk bands coalesce at the energy $E=0$, leading to the emergence of Weyl exceptional rings (which are analytically determined by Eqs.~(S7,S8) in Ref.~\cite{SMHighOrderNodalRings}). As shown in Fig.~\ref{ERings}, the non-Hermiticity  drives four higher-order Weyl nodes of the Hermitian Hamiltonian $\mathcal{H}(\mathbf{k}, \gamma=0)$ into four Weyl exceptional rings, where the real and imaginary parts of the eigenvalues vanish. These exceptional rings are protected by the $\mathcal{CP}$  symmetry, belonging to  the class $\mathcal{PC}$   \cite{PhysRevLett.123.066405}. In order to characterize their topological stability, we calculate the spectral winding number defined as \cite{PhysRevLett.123.066405} 
\begin{align}\label{berryphase22}
	\mathcal{W} = \oint_{\mathcal{L}} \frac{d\mathbf{k}}{2\pi i} \cdot \triangledown_\mathbf{k}\textrm{log det}\left[\mathcal{H}(\mathbf{k})-E(\mathbf{k}_\textrm{EP})\right],
\end{align}
where $\mathcal{L}$ is a closed path encircling one of the exceptional points on the Weyl exceptional ring [see Fig.~\ref{ERings}(a)], and $E(\mathbf{k}_\textrm{EP})=0$ is the reference energy at the corresponding exceptional point $\mathbf{k}_\textrm{EP}$. Because there exists a point gap [see Fig.~\ref{ERings}(b)] along the path  $\mathcal{L}$ for the reference point $E(\mathbf{k}_\textrm{EP})$, the spectral winding number in Eq.~(\ref{berryphase22}) is well defined, and can be nonzero due to complex eigenergies. Direct numerical calculations yield $\mathcal{W} = -1,~1,~1,~-1,$ for each nodal ring along the $z$ axis in Fig.~\ref{ERings}. The quantized non-zero winding numbers indicate that Weyl exceptional rings are topologically protected by the $\mathcal{CP}$  symmetry, and cannot be removed by small perturbations preserving symmetries.



 \begin{figure*}[!tb]
	\centering
	\includegraphics[width=17.8cm]{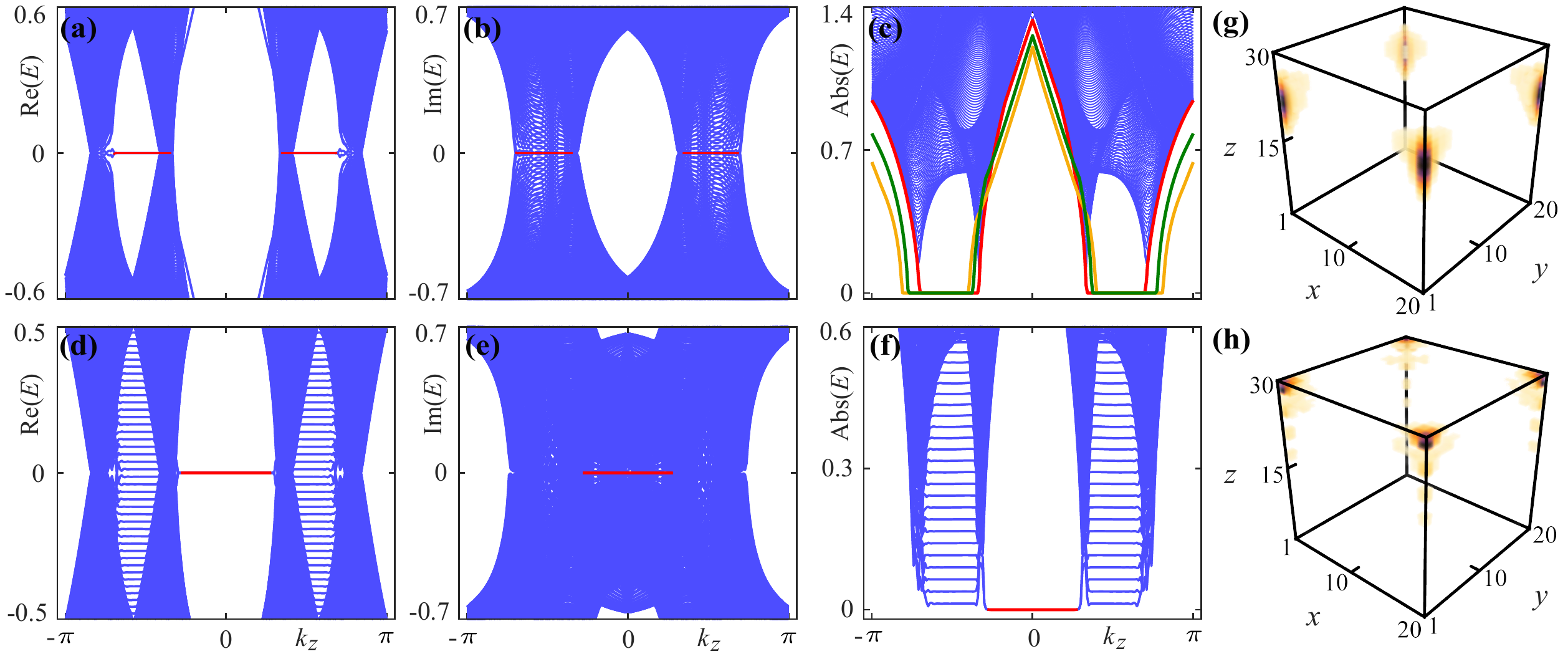}
	\caption{(a) Real, (b) imaginary, and (c) absolute values of the surface band structure along the $k_z$ direction, when the open boundary condition is imposed along the $x$ direction with 200 sites for $k_y=0$. Yellow, green and red lines in (c) denote surface states for $\gamma=0.4$, $\gamma=0.6$, and $\gamma=0.8$, respectively. Note that only the modes with zero absolute energy are surface states.  (d) Real, (e) imaginary, and (f) absolute values of the band structure of a finite-sized system with $60 \times 60$ unit cells in the $x$-$y$ plane. The red lines represent hinge states.   Probability density distributions $\abs{\psi_n(x,y,z)}^2$ of midgap modes for the eigenenergies (g) $E = 4.6\times10^{-6}(1+i)$ and (h)  $E=0.0072+0.0072i$ with open boundaries along the $x$, $y$, and $z$ directions. The number of unit cells is 20 $\times$ 20 $\times$  30. The parameters used here are: $m_0 = 1.5$, $m_1  = -1$, $v_z=0.8$, $\gamma=0.8$, and $\Delta_0 =0.8$.}\label{surfacehinge}
\end{figure*}

\textit{Surface bands}.---Since Weyl exceptional rings are transformed from the Weyl nodes by a non-Hermitian perturbation,  they can also carry topological charges of the Berry flux characterized by the first Chern number, which is defined on the closed surface $\mathcal{S}$ [see Fig.~\ref{ERings}(c)] as
\begin{equation}\label{chern}
	\mathcal{C} = \frac{1}{2\pi i}\sum_n \oint_\mathcal{S}  \!\!  \grad_{\mathbf{k}} \times \bra{\tilde{u}_n(\mathbf{k})} \grad_{\mathbf{k}} \ket{u_n(\mathbf{k})}\cdot d\mathbf{S},
\end{equation}
where $n$ is taken over the occupied bands, and $\ket{u_n(\mathbf{k})}$ ($\ket{\tilde{u}_n(\mathbf{k})}$) being the left (right) eigenvector for the $n$th band. The numerical calculations show that four Weyl exceptional rings carry topological charges with $\mathcal{C} = -1,~1,~1,~-1,$ for each ring along the $k_z$ direction, when each Weyl exceptional ring is enclosed by the surface $\mathcal{S}$. Otherwise,  $\mathcal{C} = 0$ when each ring is located outside the surface $\mathcal{S}$.

The non-zero Chern numbers indicate that there exist surface states, which connect two Weyl exceptional rings with opposite Chern numbers under the open boundary condition. Figure \ref{surfacehinge}(a-c) shows the real, imaginary and absolute parts of surface-band spectra when the open boundary is imposed along the $x$ direction. The non-Hermitian topological semimetal $\mathcal{H}(\mathbf{k})$ has zero-energy surface states. These surface states are bounded by the projection of Weyl exceptional rings onto the $k_z$ axis. As the non-Hermitian term $\gamma$ varies, the bounded range of zero-energy surface states changes, as shown in Fig.~\ref{surfacehinge}(c). Therefore, the non-Hermitian Weyl-exceptional-ring semimetal $\mathcal{H}(\mathbf{k})$ has the features of first-order topological semimetals.

\textit{Hinge bands}.---We now proceed to investigate its higher-order topological phase in Eq.~(\ref{H1}). We consider to impose the open boundaries along both the $x$ and $y$ directions, and figure \ref{surfacehinge}(d-f) presents the real, imaginary and absolute parts of hinge-band spectra. Remarkably, there exist hinge Fermi arcs, which connect with the projection of two Weyl exceptional rings closest to $k_z=0$ onto the hinges. This indicates that the non-Hermitian band coalescences in zero-energy bulk bands lead to the HOWERS. Therefore, the Weyl exceptional rings in the non-Hermitian semimetal $\mathcal{H}(\mathbf{k})$ have both the first-order and higher-order topological features. In Fig.~\ref{surfacehinge}(g,h), we present the probability density distributions of two arbitrary mid-gap states for open boundary conditions along the $x$, $y$ and $z$ directions. In contrast to Hermitian Weyl semimetals, the hinge Fermi-arc states studied here show non-Hermitian skin effects, and thus hinge modes are localized towards corners.  



\textit{Effective boundary theory}.---
For an intuitive understanding of the HOWERS and the emergence of hinge Fermi-arc states, we develop an effective boundary theory to derive the low-energy surface-state Hamiltonian in the gapped bulk-band regime for the relatively small $\gamma$ and $\Delta_0$ (See the details in Ref.~\cite{SMHighOrderNodalRings}). We label the four surfaces of a cubic sample as $\textrm{\Rmnum{1}}, \textrm{\Rmnum{2}}, \textrm{\Rmnum{3}}, \textrm{\Rmnum{4}}$, corresponding to the surface states localized at $x=1$, $y=1$, $x=L$, and $y=L$.  We firstly consider the system under the open boundary condition along the $x$ direction, and periodic boundary conditions along both the $y$ and $z$ directions.  After a partial Fourier transformation along the $k_x$ direction, the Hamiltonian $\mathcal{H}(\mathbf{k})$ in Eq.~(\ref{H1}) becomes  $\mathcal{H}_x = \mathcal{H}_{0} + \mathcal{H}_{1}$, with
\begin{align}\label{Hx0}
	\mathcal{H}_{0} = & \sum_{x,k_y,k_z}\left(\Psi_{x,k_y,k_z}^\dagger T \Psi_{x+1,k_y,k_z}+ \textrm{H.c.}\right) \nonumber \\
	& +  \sum_{x,k_y,k_z} \Psi_{x,k_y,k_z}^\dagger   M \Psi_{x,k_y,k_z},
\end{align}
here $T$ and $M$ are $T=-\frac{1}{2} s_z \sigma_z   - \frac{i}{2}   s_x \sigma_z$ and $M=\left(m_0 - \cos k_y + m_1 \cos k_z \right) s_z \sigma_z    + \left(v_z \sin k_z  + i \gamma\right) s_z$,  and 
\begin{align}\label{Hx1}
	\mathcal{H}_{1} = &\sum_{x,k_y,k_z} \Psi_{x,k_y,k_z}^\dagger \left( \sin k_y s_y \sigma_z  - \Delta_0 \cos k_y  \sigma_x \right) \Psi_{x,k_y,k_z} \nonumber \\
	& +   \sum_{x,k_y,k_z} \left(\Psi_{x,k_y,k_z}^\dagger \frac{\Delta_0}{2}\sigma_x \Psi_{x+1,k_y,k_z}+ \textrm{H.c.}\right),
\end{align}
where $x$ is the integer-valued coordinate taking values from $1$ to $L$, and $\Psi_{x,k_y,k_z}^\dagger$ creates a fermion with spin and orbital degrees of freedom on site $x$ and momentum $k_y$ and $k_z$. By assuming a small $\Delta_0$, and taking $k_y$ to be close to 0, $\mathcal{H}_{1}$ is treated as a perturbation. 

Since the Hamiltonian $\mathcal{H}_{0}$ in Eq.~(\ref{Hx0}) is non-Hermitian, we calculate its left and right eigenstates. We first  solve the right eigenstates. In order to solve the surface states localized at the boundary $x=1$, we choose a trial solution $\psi_R(x) = \lambda_R^{x} \phi_R$, where $\lambda_R$ is a parameter determining the localization length with $\abs{\lambda_R}<1$, and $\phi_R$ is a four-component vector. Plugging this trial solution into Eq.~(\ref{Hx0}) for $k_y=0$, we have the following eigenvalue equations:   
\begin{align}\label{s1}
	\left(\lambda_R^{-1} T^\dagger + M + \lambda_R T\right) \phi_R= E \phi_R, ~~~~~\textrm{in the bulk},
\end{align}
and
\begin{align}\label{s2}
	\left(M + \lambda_R T\right) \phi_R= E\phi_R, ~~~~~  \textrm{at the boundary} ~x=1.
\end{align}
By considering the semi-infinite limit $L \to \infty$, and requiring the states to have the same eigenenergies in the bulk and at the boundary, we have $\lambda_R^{-1} T^\dagger \phi_R = 0$. This leads to $E=0$, and two eigenvectors with $\phi_{1,R} = (-i,~0,~1,~0)^T$, and $\phi_{2,R} =  (0,~-i,~0,~1)^T$. The corresponding localization parameters are $\lambda_{1,R} = 1 - m_0 - m_1 \cos k_z - v_z \sin k_z  - i \gamma$, and $\lambda_{2,R} = 1 - m_0 - m_1 \cos k_z + v_z \sin k_z  + i \gamma$, respectively. 

For the surface states localized at the boundary $x=1$, we require $\abs{\lambda_R}<1$, then 
\begin{align}\label{eeqq}
\left[\left(1 - m_0 - m_1 \cos k_z \pm v_z \sin k_z\right)^2 + \gamma^2\right]^{1/2} < 1.
\end{align}
According to Eq.~(\ref{eeqq}), as $k_z$ increases from $0$ to $\pi$ (or decreases from $0$ to $-\pi$), the non-Hermitian system $\mathcal{H}_{0}$ first supports two surface states localized at the boundary $x=1$, and then only one surface state as $\abs{k_z}$ exceeds a critical value (i.e., one of the exceptional points at which phase transition takes place). As shown in Fig.~\ref{lambda}, two surface states exist only in a finite region of $k_z$ inbetween two exceptional rings closest to $k_z=0$ for   small $\gamma$. A surface energy gap, or a mass term, can exist only when two surface eigenstates coexist. Thus, the hinge states, regarded as boundary states between domains of opposite masses, appear only in a finite range of $k_z$.

\begin{figure}[!tb]
	\centering
	\includegraphics[width=5.0cm]{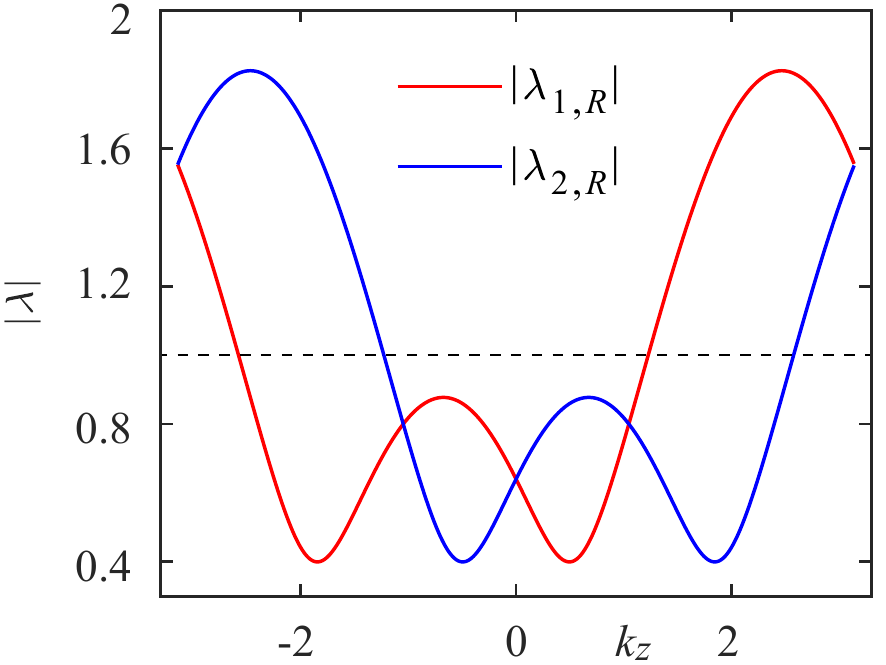}
	\caption{$\abs{\lambda_{1, R}}$ and $\abs{\lambda_{2, R}}$ versus $k_z$, according to Eq.~(\ref{eeqq}), with $m_0 = 1.5$, $m_1  = -1$, $v_z=0.8$, and $\gamma=0.4$. The dashed black line denotes $\abs{\lambda} = 1$, which is guided for the eyes.}\label{lambda}
\end{figure}

The left eigenstates $\psi_L(x)$ can be obtained by using the same procedure for deriving right eigenstates \cite{SMHighOrderNodalRings}. Therefore, considering the $k_z$ region where the system supports two surface states, and projecting the Hamiltonian $\mathcal{H}_{1}$ in Eq.~(\ref{Hx1}) into the subspace spanned by the above left and right eigenstates as $\mathcal{H}_{\textrm{surf},\alpha\beta}^\textrm{\Rmnum{1}} = \psi_{\alpha,L}^* \mathcal{H}_{1} \psi_{\beta,R}$, we obtain the effective boundary Hamiltonian at the surface $\textrm{\Rmnum{1}}$ as
\begin{align}\label{Surface1}
	\mathcal{H}_{\textrm{surf,x}}^\textrm{\Rmnum{1}}(k_y, k_z) =   k_y \sigma_z - \mu \sigma_x,
\end{align}
with
\begin{align}\label{Surface12}
\mu = \frac{\Delta_0 \mathcal{N}_1 \mathcal{N}_2 \lambda_{1,R}\lambda_{2,R}
	\left(2-\lambda_{1,R}-\lambda_{2,R}\right)}{\left(1-\lambda_{1,R}\lambda_{2,R}\right)},
\end{align}
where $\mathcal{N}_1$ and $\mathcal{N}_2$ are normalized parameters \cite{SMHighOrderNodalRings}, and we have ignored the terms of order higher than $k_y$.

Considering the same procedure above, we can have the boundary states localized at surfaces $\textrm{\Rmnum{2}}$, $\textrm{\Rmnum{3}}$ and $\textrm{\Rmnum{4}}$ as
\begin{align}\label{Surfacey2}
	\mathcal{H}_{\textrm{surf,y}}^\textrm{\Rmnum{2}}(k_x, k_z) =   -k_x \sigma_z + \mu \sigma_x,
\end{align}
\begin{align}\label{Surface3}
	\mathcal{H}_{\textrm{surf,x}}^\textrm{\Rmnum{3}}(k_y, k_z) =   -k_y \sigma_z - \mu \sigma_x,
\end{align}
\begin{align}\label{Surface4}
	\mathcal{H}_{\textrm{surf,y}}^\textrm{\Rmnum{4}}(k_x, k_z) =   k_x \sigma_z + \mu \sigma_x.
\end{align}

According to the surface Hamiltonians in Eqs.~(\ref{Surface1})-(\ref{Surface4}), for each $k_z$, the boundary states show the same coefficients for the kinetic energy terms, but mass terms on two neighboring boundaries always have opposite signs. Therefore, mass domain walls appear at the intersection of two neighboring boundaries, and these two boundaries can share a common zero-energy boundary state (analogous to the Jackiw-Rebbi zero modes \cite{PhysRevD.13.3398}) in spite of complex-valued $\mu$, which corresponds to the hinge Fermi-arc states at each $k_z$.  Moreover, these hinge Fermi-arc states exist only in a finite $k_z$ region limited by the condition in Eq.~(\ref{eeqq}). This explains why the Hamiltonian $\mathcal{H}(\mathbf{k})$ shows both first-order and higher-order topological features for small $\gamma$.

\begin{figure}[!tb]
	\centering
	\includegraphics[width=8.4cm]{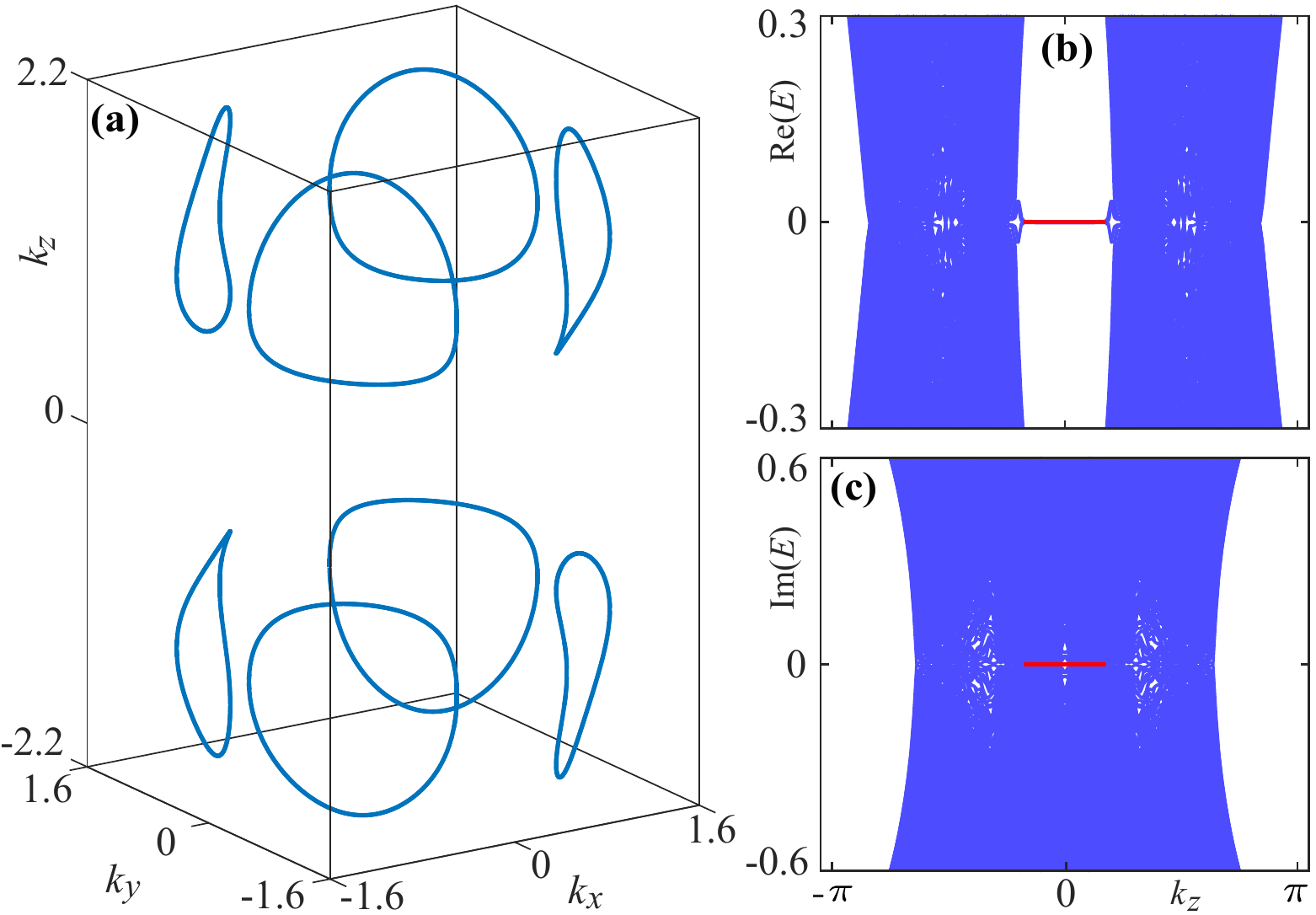}
	\caption{  (a) Weyl exceptional rings (blue curves) for $\gamma=1.1$.  The bulk bands of the non-Hermitian Hamiltonian  exhibit eight Weyl exceptional rings. (b) Real and (c) imaginary parts of the band structure of a finite-sized system with $80 \times 80$ unit cells in the $x$-$y$ plane. The red lines represent hinge states.   The parameters used here are: $m_0 = 1.5$, $m_1  = -1$, $v_z=0.8$, and $\Delta_0 =0.8$.}\label{SecondRings}
\end{figure}

\textit{Topological phase transitions}.---The dissipative term $\gamma$ in Eq.~(\ref{H1}) can induce topological phase transitions. For small $\gamma$, $\mathcal{H}(\mathbf{k})$ exhibits four exceptional rings, as discussed above. As $\gamma$ increases, two Weyl exceptional rings with opposite Chern numbers, located in the positive (negative) $k_z$ axis  [see Fig.~\ref{ERings}(a)], move towards each other. At the critical value of $\gamma \simeq 1$, a topological phase transition occurs, where two Weyl exceptional rings carrying opposite topological charges are coupled and annihilated. Then the system evolves into a new topological semimetal with eight Weyl exceptional rings,  as shown in Fig.~\ref{SecondRings}(a). These rings are topologically stable due to the non-zero spectral winding numbers defined in Eq.~(\ref{berryphase22}). However, the Chern number, defined in Eq.~(\ref{chern}), is zero when the closed surface $\mathcal{S}$ encloses four exceptional rings located in the positive (negative) $k_z$ axis. Therefore, there exist no surface Fermi-arc states when the boundary is open along the $x$ or $y$ direction. To further check the higher-order topological phases, figure \ref{SecondRings}(b,c) shows the band structures for a finite-sized system in the $x$-$y$ plane for $\gamma=1.1$. The semimetal supports in-gap hinge Fermi-arc states, indicating that the Weyl-exceptional-ring semimetal has only higher-order topological features in the strong dissipation regime.  
	

\textit{Conclusion}.---We have proposed a theoretical model to realize non-Hermitian higher-order topological semimetals, and identify a new type of bulk-band degeneracies, i.e., Weyl exceptional rings in higher-order topological phases. These rings are characterized by the spectral winding number and Chern number. Remarkably,  non-Hermitian higher-order topological semimetals, in the presence of Weyl exceptional rings, show the coexistence of surface and hinge Fermi arcs. Moreover, the dissipative terms can cause the coupling of two exceptional rings with opposite topological charges, so as to induce topological phase transitions.  Non-Hermitian higher-order Weyl semimetals have not been explored in the past, and these studies would advance the development of this field. 

Noted added: After this work was submitted, we became aware of a related work discussing non-Hermitian higher-order Weyl semimetals  with different focus \cite{arXiv:2107.00024}.

\begin{acknowledgments}
T.L. acknowledges the support from the Startup Grant of South China University of Technology (Grant No.~20210012). F.N. is supported in part by:
Nippon Telegraph and Telephone Corporation (NTT) Research,
the Japan Science and Technology Agency (JST) [via
the Moonshot RD Grant Number JPMJMS2061, and
the Centers of Research Excellence in Science and Technology (CREST) Grant No.~JPMJCR1676],
the Japan Society for the Promotion of Science (JSPS)
[via the Grants-in-Aid for Scientific Research (KAKENHI) Grant No.~JP20H00134 and the
JSPS–RFBR Grant No.~JPJSBP120194828],
the Army Research Office (ARO) (Grant No.~W911NF-18-1-0358),
the Asian Office of Aerospace Research and Development (AOARD) (via Grant No.~FA2386-20-1-4069), and
the Foundational Questions Institute Fund (FQXi) via Grant No.~FQXi-IAF19-06.
\end{acknowledgments}

%

\clearpage \widetext
\begin{center}
	\section{Supplemental Material for ``Higher-Order Weyl-Exceptional-Ring Semimetals"}
\end{center}
\setcounter{equation}{0} \setcounter{figure}{0}
\setcounter{table}{0} \setcounter{page}{1} \setcounter{secnumdepth}{3} \makeatletter
\renewcommand{\theequation}{S\arabic{equation}}
\renewcommand{\thefigure}{S\arabic{figure}}
\renewcommand{\bibnumfmt}[1]{[S#1]}
\renewcommand{\citenumfont}[1]{S#1}

\makeatletter
\def\@hangfrom@section#1#2#3{\@hangfrom{#1#2#3}}
\makeatother


\section{Hermitian Higher-Order Weyl Semimetals}

As shown in the main text, we consider the following minimal Hamiltonian
\begin{align}\label{H1}
	\mathcal{H}(\mathbf{k}) = ~&\left(m_0 - \cos k_x  - \cos k_y + m_1 \cos k_z \right) s_z \sigma_z    + \left(v_z \sin k_z  + i \gamma\right) s_z +  \sin k_x   s_x \sigma_z  + \sin k_y s_y \sigma_z  \nonumber \\ 
	& + \Delta_0  \left(\cos k_x - \cos k_y \right) \sigma_x,
\end{align}
In the absence of the non-Hermitian term (i.e., $\gamma=0$), the Hermitian Hamiltonian $\mathcal{H}(\mathbf{k}, \gamma=0)$ breaks the inversion symmetry $\mathcal{P}$,  but preservers the time-reversal symmetry $\mathcal{T} = i s_x \sigma_x \mathcal{K}$, with $\mathcal{K}$ being the complex conjugation operator. The eigenenergy $\mathcal{E}$ of the Hamiltonian for $\gamma=0$ is
\begin{align}\label{E1}
	\mathcal{E}^2 = \left(\left\lvert v_z \sin k_z\right \rvert \pm \sqrt{\left(m_0-\cos k_x - \cos k_y +m_1 \cos k_z \right)^2+\Delta ^2 \left(\cos k_x-\cos k_y\right)^2}\right)^2+\sin^2 k_x +\sin^2 k_y.
\end{align}
According to Eq.~(\ref{E1}), the Hermitian Hamiltonian supports higher-order Weyl nodes located at   $(k_x,~k_y,~k_z) = (0,~0,~k_w)$, where  $k_w$ satisfies
\begin{align}\label{kw}
	v_z^2 \sin^2  k_w \pm \left(m_0  -2+m_1\cos k_w \right)^2 = 0.
\end{align}

\begin{figure}[!htb]
	\centering
	\includegraphics[width=18cm]{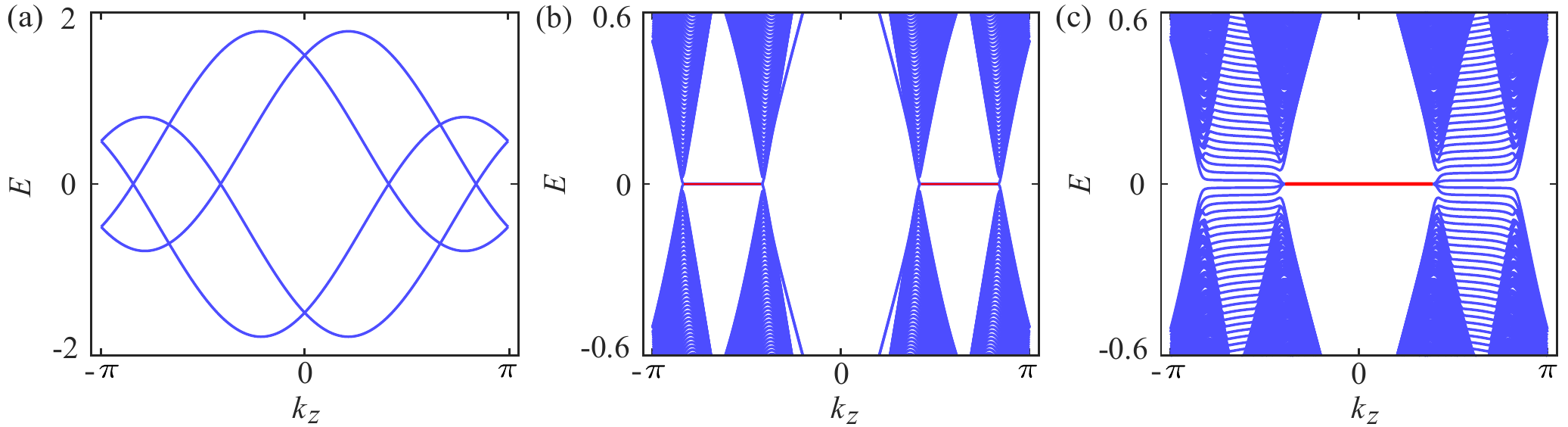}
	\caption{(a) Bulk band structure along the $k_z$ direction for $k_x=k_y=0$. There exist four Weyl nodes, at which bulk bands are two-fold degenerate and  eigenenergies are zero. (b) Surface band structure along the $k_z$ direction under the open boundary condition along the $x$ direction for $k_y=0$. Two Weyl nodes located at the negative (positive) $k_z$ axis are connected by surface Fermi arcs (red lines). (c) Band structure of a finite-sized system with $60 \times 60$ unit cells in the $x$-$y$ plane. The hinge Fermi arcs (red lines) connect two Weyl nodes closest to $k_z = 0$, which are second-order Weyl nodes.  The parameters used here are:  $m_0 = 1.5$, $m_1  = -1$, $v_z=0.8$, $\gamma=0$, and $\Delta_0 =0.8$.}\label{FigSM1}
\end{figure}

As shown in Fig.~\ref{FigSM1}(a), there exist four Weyl nodes in momentum space, which are connected through surface Fermi arcs [see Fig.~\ref{FigSM1}(b)] for the open boundary condition along the $x$ direction. Moreover, when the boundaries along both the $x$ and $y$ directions are opened, hinge Fermi-arc states appear, which connect the two Weyl nodes closest to $k_z = 0$.  Therefore, the Hermitian Hamiltonian $\mathcal{H}(\mathbf{k}, \gamma=0)$ is a hybrid-order Weyl semimetal.

%
%


\section{Weyl Exceptional Rings}

In the presence of the non-Hermitian term, the eigenenergy $E$ of the Hamiltonian $\mathcal{H}(\mathbf{k})$ can be written as
%
\begin{align}\label{Eergy2}
	E^2 = &\left(m_0- \cos k_x-\cos k_y+m_1 \cos k_z\right)^2+\Delta^2 \left(\cos k_x-\cos k_y\right)^2-\left(\gamma -i v_z \sin k_z\right)^2 +\sin^2 k_x + \sin^2 k_y \nonumber \\
	& \pm 2 \left(i \gamma + v_z \sin k_z\right) \left[\left(m_0- \cos k_x-\cos k_y+m_1 \cos k_z\right)^2+\Delta^2 \left(\cos k_x-\cos k_y\right)^2\right]^{1/2}
\end{align}
%

To have the bands coalescence, we require $E^2 = 0$, namely,
\begin{align}\label{e1}
	\left(m_0- \cos k_x-\cos k_y+m_1 \cos k_z\right)^2+\Delta^2 \left(\cos k_x-\cos k_y\right)^2 = v_z^2 \sin^2 k_z,
\end{align}
\begin{align}\label{e2}
	\left(\left\lvert v_z \sin k_z\right \rvert - \sqrt{\left(m_0- \cos k_x - \cos k_y +m_1 \cos k_z \right)^2+\Delta ^2 \left(\cos k_x-\cos k_y\right)^2}\right)^2+\sin^2 k_x +\sin^2 k_y = \gamma^2.
\end{align}
In the above, without loss of generality, we have required $v_z > 0$. According to Eqs.~(\ref{e1}) and (\ref{e2}), we have
\begin{align}\label{e3}
	\sin^2 k_x +\sin^2 k_y = \gamma^2, 
\end{align}
and
\begin{align}\label{e4}
	\left(m_0- \cos k_x-\cos k_y+m_1 \cos k_z\right)^2+\Delta^2 \left(\cos k_x-\cos k_y\right)^2 = v_z^2 \sin^2 k_z.
\end{align}

\begin{figure}[!tb]
	\centering
	\includegraphics[width=18.0cm]{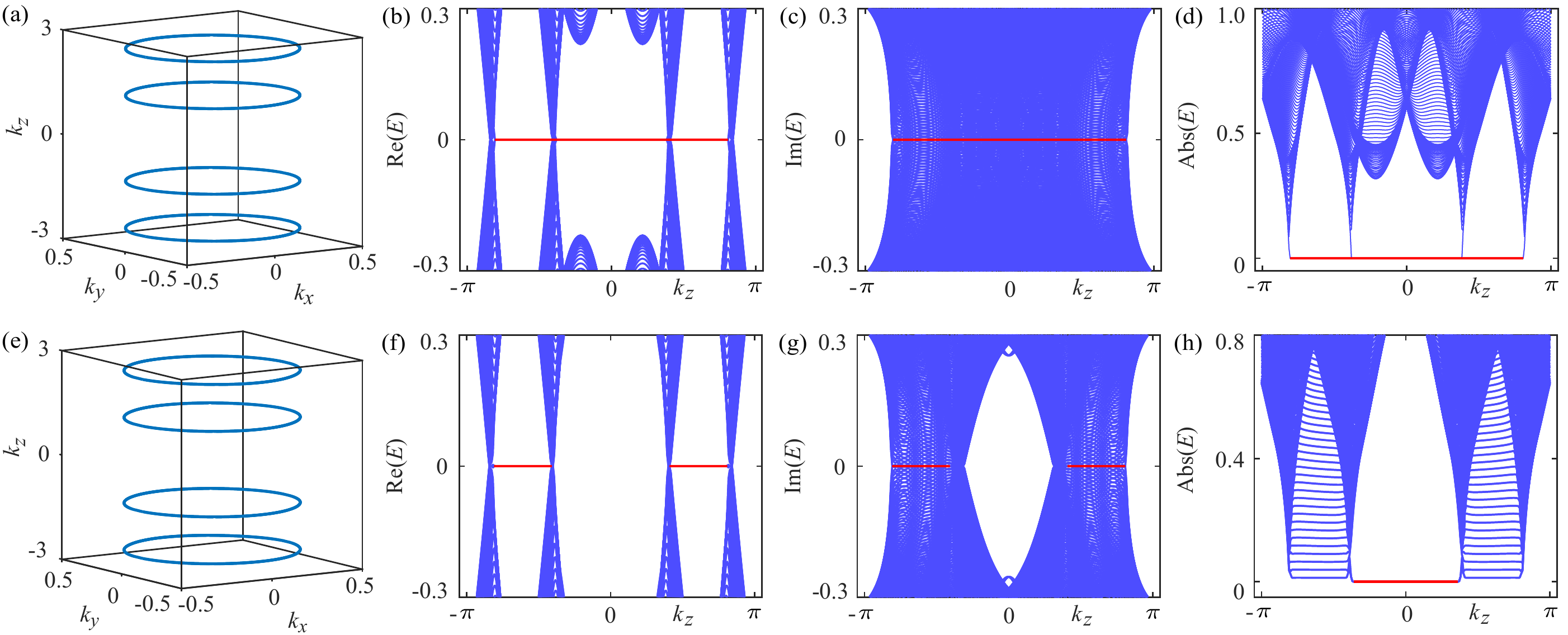}
	\caption{First-order (first row) and higher-order (second row) topological semimetals for $\gamma=0.4$. (a) Four Weyl exceptional rings along the $k_z$ direction in the first-order topological  semimetal for $\Delta_0 =0$. (b) Real, (c) imaginary, and (d) absolute values of the surface band structure along the $k_z$ direction for $\Delta_0 =0$, when the open boundary condition is imposed along the $x$ direction with 200 sites for $k_y=0$. Note that only the modes with zero absolute energy are surface states (red lines). (e) Four Weyl exceptional rings along the $k_z$ direction in the second-order topological semimetal for $\Delta_0 =0.8$. (f) Real and (g) imaginary parts of the surface band structure along the $k_z$ direction for $\Delta_0 =0.8$, when the open boundary condition is imposed along the $x$ direction with 200 sites for $k_y=0$. (h) Absolute values of the band structure of a finite-sized system with $60 \times 60$ unit cells in the $x$-$y$ plane. Note that only the modes with zero absolute energy are surface and hinge states (red lines). The common parameters used here are: $m_0 = 1.5$, $m_1  = -1$, $v_z=0.8$, and $\gamma=0.4$.}\label{surfaceSm}
\end{figure}

\section{Effective surface Hamiltonian in the gapped regimes}

\begin{figure}[!tb]
	\centering
	\includegraphics[width=8.0cm]{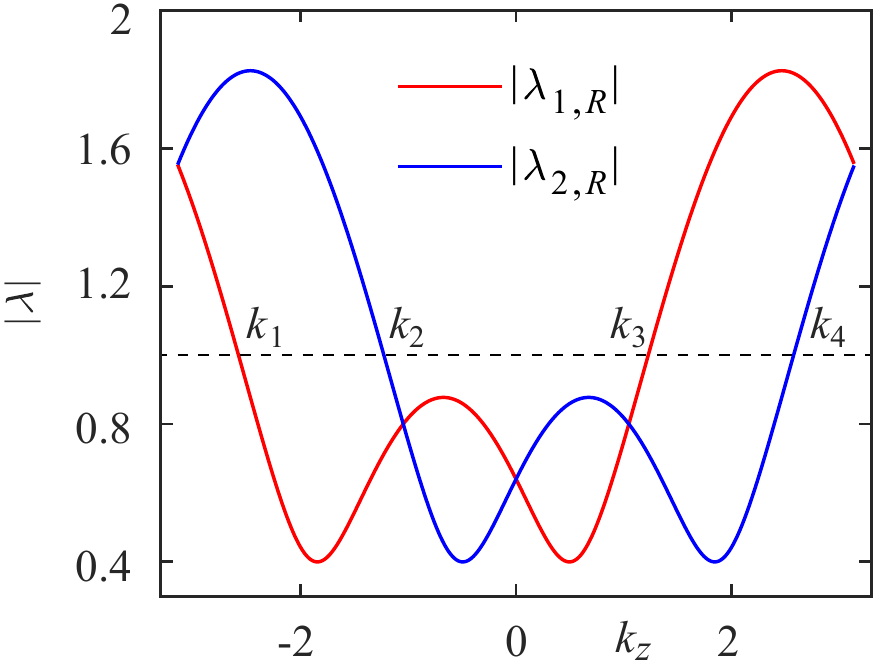}
	\caption{$\abs{\lambda_{1, R}}$ and $\abs{\lambda_{2, R}}$ versus $k_z$, according to Eq.~(\ref{eeqq}), with $m_0 = 1.5$, $m_1  = -1$, $v_z=0.8$, and $\gamma=0.4$. The horizontal dashed black line denotes $\abs{\lambda} = 1$, which is just guided for  eyes. The projections of four exceptional rings of $\mathcal{H}_{0}$ are located at $k_1$, $k_2$, $k_3$ and $k_4$. As $k_z$ increases from $0$ to $\pi$ (or decreases from $0$ to $-\pi$), the non-Hermitian system $\mathcal{H}_{0}$ first supports two surface states localized for $k_2<k_z<k_3$ at the boundary $x=1$, and then only one surface state localized for $k_1<k_z<k_2$ or $k_3<k_z<k_4$ at the boundary as $\abs{k_z}$ exceeds a critical value (i.e., $k_2$ and $k_3$). }\label{lambda}
\end{figure}

For $\Delta_0=0$, the Hamiltonian $\mathcal{H}(\mathbf{k})$ in Eq.~(\ref{H1}) is a first-order topological semimetal with the Weyl exceptional rings [see Fig.~\ref{surfaceSm}(a-d)], while the $\Delta_0$ leads to a higher-order Weyl-exceptional-ring semimetal [see Fig.~\ref{surfaceSm}(e-h)]. Thus, the $\Delta_0$ term gaps out the surface bands in the finite $k_z$ region in the first Brillouin zone. In this part, we derive the low-energy effective Hamiltonians of surface bands in the gapped bulk-band regime for the relatively small $\gamma$ and $\Delta_0$. We label the four surfaces of a cubic sample as $\textrm{\Rmnum{1}}, \textrm{\Rmnum{2}}, \textrm{\Rmnum{3}}, \textrm{\Rmnum{4}}$, corresponding to the boundary states localized at $x=1$, $y=1$, $x=L$, and $y=L$. 

We first consider the system under  open boundary condition along the $x$ direction, and periodic boundary conditions along both the $y$ and $z$ directions.  After a partial Fourier transformation along the $k_x$ direction, the Hamiltonian $\mathcal{H}(\mathbf{k})$ in Eq.~(\ref{H1}) becomes
\begin{align}\label{Hx}
	\mathcal{H}_x(k_y, ~k_z) = ~&\sum_{x,k_y,k_z} \Psi_{x,k_y,k_z}^\dagger \left[\left(m_0 - \cos k_y + m_1 \cos k_z \right) s_z \sigma_z    + \left(v_z \sin k_z  + i \gamma\right) s_z  + \sin k_y s_y \sigma_z  \right.  \nonumber \\
	& \left. - \Delta_0 \cos k_y  \sigma_x \right] \Psi_{x,k_y,k_z} +  \sum_{x,k_y,k_z} \left[\Psi_{x,k_y,k_z}^\dagger \left(-\frac{1}{2} s_z \sigma_z   - \frac{i}{2}   s_x \sigma_z  + \frac{\Delta_0}{2} \sigma_x\right)     \Psi_{x+1,k_y,k_z}+ \textrm{H.c.}\right] ,
\end{align}
where $x$ is the integer-valued coordinate taking values from $1$ to $L$, and $\Psi_{x,k_y,k_z}^\dagger$ creates a fermion with spin and orbital degrees of freedom on site $x$ and momentum $k_y$ and $k_z$. By assuming a small $\Delta_0$ and taking $k_y$ to be close to 0, we rewrite $\mathcal{H}_x$  as $\mathcal{H}_x = \mathcal{H}_{0} + \mathcal{H}_{1}$, with
\begin{align}\label{Hx0}
	\mathcal{H}_{0} = \sum_{x,k_y,k_z} \Psi_{x,k_y,k_z}^\dagger   M \Psi_{x,k_y,k_z}  +  \sum_{x,k_y,k_z} \left(\Psi_{x,k_y,k_z}^\dagger T \Psi_{x+1,k_y,k_z}+ \textrm{H.c.}\right) ,
\end{align}
where $M=\left(m_0 - \cos k_y + m_1 \cos k_z \right) s_z \sigma_z    + \left(v_z \sin k_z  + i \gamma\right) s_z$, and $T=-\frac{1}{2} s_z \sigma_z   - \frac{i}{2}   s_x \sigma_z$, and 
\begin{align}\label{Hx1}
	\mathcal{H}_{1} = ~\sum_{x,k_y,k_z} \Psi_{x,k_y,k_z}^\dagger \left( \sin k_y s_y \sigma_z  - \Delta_0 \cos k_y  \sigma_x \right) \Psi_{x,k_y,k_z} +   \sum_{x,k_y,k_z} \left(\Psi_{x,k_y,k_z}^\dagger \frac{\Delta_0}{2}\sigma_x \Psi_{x+1,k_y,k_z}+ \textrm{H.c.}\right),
\end{align}
where $\mathcal{H}_{1}$ is treated as a perturbation. 

Since the Hamiltonian $\mathcal{H}_{0}$ in Eq.~(\ref{Hx0}) is non-Hermitian, we calculate its left and right eigenstates. We first solve the right eigenstates. In order to solve the surface states localized at the boundary $x=1$, we choose a trial solution $\psi_R(x) = \lambda_R^{x} \phi_R$, where $\lambda_R$ is a parameter determining the localization length with $\abs{\lambda_R}<1$, and $\phi_R$ is a four-component vector. Plugging this trial solution into Hamiltonian $\mathcal{H}_{0}$ in Eq.~(\ref{Hx0}) for $k_y=0$, we have the following eigenvalue equations:   
\begin{align}\label{s1}
	\left(\lambda_R^{-1} T^\dagger + M + \lambda_R T\right) \phi_R= E \phi_R, ~~~~~\textrm{in the bulk},
\end{align}
and
\begin{align}\label{s2}
	\left(M + \lambda_R T\right) \phi_R= E\phi_R, ~~~~~  \textrm{at the boundary} ~x=1.
\end{align}
By considering the semi-infinite limit $L \to \infty$, and requiring the states have the same eigenenergy in the bulk and at the boundary, we have $\lambda_R^{-1} T^\dagger \phi_R = 0$. This leads to $E=0$, and two corresponding eigenstates $\psi_{1,R}$  and $\psi_{2,R}$.  The eigenstate $\psi_{1,R}$ is written as 
\begin{align}\label{eig1}
	\psi_{1,R} = \mathcal{N}_1 (\lambda_{1,R} \phi_{1,R},~\lambda_{1,R}^2 \phi_{1,R},~\lambda_{1,R}^3 \phi_{1,R},~\dots),
\end{align}
with
\begin{align}\label{eig12}
	\phi_{1,R} = (-i,~0,~1,~0)^T, ~~~\textrm{and}~~ \lambda_{1,R} =  1 - m_0 - m_1 \cos k_z - v_z \sin k_z  -i \gamma.
\end{align}
The eigenstate $\psi_{2,R}$ is
\begin{align}\label{eig2}
	\psi_{2,R} = \mathcal{N}_2 (\lambda_{2,R} \phi_{2,R},~\lambda_{2,R}^2 \phi_{2,R},~\lambda_{2,R}^3 \phi_{2,R},~\dots),
\end{align}
with
\begin{align}\label{eig22}
	\phi_{2,R} =  (0,~-i,~0,~1)^T, ~~\textrm{and}~~ \lambda_{2,R} =  1 - m_0 - m_1 \cos k_z + v_z \sin k_z  + i \gamma.
\end{align}

For the surface states localized at the boundary $x=1$, we require  $\abs{\lambda_{1,R}}<1$ and $\abs{\lambda_{2,R}}<1$, then we have 
\begin{align}\label{eeqq}
	\left[\left(1 - m_0 - m_1 \cos k_z - v_z \sin k_z\right)^2 + \gamma^2\right]^{1/2} < 1, ~~\textrm{and}~~~  \left[\left(1 - m_0 - m_1 \cos k_z + v_z \sin k_z\right)^2 + \gamma^2\right]^{1/2} 	  < 1.
\end{align}
According to Eq.~(\ref{eeqq}), as $k_z$ increases from $0$ to $\pi$ (or decreases from $0$ to $-\pi$), the non-Hermitian system $\mathcal{H}_{0}$ first supports two surface states localized at the boundary $x=1$, and then only one surface state as $\abs{k_z}$ exceeds a critical value (i.e., one of exceptional points at which a phase transition takes place). As shown in Fig.~\ref{lambda}, two surface states exist only in a finite region of $k_z$ inbetween two exceptional rings closest to $k_z=0$ for   small $\gamma$. A surface energy gap, or a mass term, can exist only when two surface eigenstates coexist. Thus, the hinge states, regarded as boundary states between domains of opposite masses, appear only in a finite range of $k_z$.

We now proceed to solve the left eigenstates with a trial solution $\psi_L(x) = \lambda_L^{x} \phi_L$. As the same procedure for deriving the right eigenstates, we have the following eigenvalue equations:   
\begin{align}\label{s1L}
	\left(\lambda_L^{-1} T^\dagger + M^\dagger + \lambda_L T\right) \phi_L= E \phi_L, ~~~~~\textrm{in the bulk},
\end{align}
and
\begin{align}\label{s2L}
	\left(M^\dagger + \lambda_L T\right) \phi_L= E\phi_L, ~~~~~  \textrm{at the boundary} ~x=1.
\end{align}
By considering the semi-infinite limit, we have $E=0$, and two corresponding eigenstates $\psi_{1,L}$  and $\psi_{2,L}$.  The eigenstate $\psi_{1,L}$ is written as 
\begin{align}\label{eig1L}
	\psi_{1,L} = \mathcal{N}_1^* (\lambda_{1,L} \phi_{1,L},~\lambda_{1,L}^2 \phi_{1,L},~\lambda_{1,L}^3 \phi_{1,L},~\dots),
\end{align}
with
\begin{align}\label{eig12L}
	\phi_{1,L} = (-i,~0,~1,~0)^T, ~~~\textrm{and}~~ \lambda_{1,L} =  1 - m_0 - m_1 \cos k_z - v_z \sin k_z  +i \gamma.
\end{align}
The eigenstate $\psi_{2,L}$ is
\begin{align}\label{eig2L}
	\psi_{2,L} = \mathcal{N}_2^* (\lambda_{2,L} \phi_{2,L},~\lambda_{2,L}^2 \phi_{2,L},~\lambda_{2,L}^3 \phi_{2,L},~\dots),
\end{align}
with
\begin{align}\label{eig22L}
	\phi_{2,L} =  (0,~-i,~0,~1)^T, ~~\textrm{and}~~ \lambda_{2,L} =  1 - m_0 - m_1 \cos k_z + v_z \sin k_z  - i \gamma.
\end{align}
In Eqs.~(\ref{eig1}, ~\ref{eig2}, ~\ref{eig1L},~\ref{eig2L}), the constants $\mathcal{N}_1$ and $\mathcal{N}_2$ are solved by biorthogonal conditions as 
\begin{align}\label{bior1}
	\mathcal{N}_1= \sqrt{\left(1- \lambda_{1, L}^* \lambda_{1, R}\right)/\left(2 \lambda_{1, L}^* \lambda_{1, R} \right)},
\end{align}
\begin{align}\label{bior2}
	\mathcal{N}_2= \sqrt{\left(1- \lambda_{1, L}^* \lambda_{1, R}\right)/\left(2 \lambda_{1, L}^* \lambda_{1, R} \right)}.
\end{align}

For the $k_z$ region where the system supports two surface states, projecting the Hamiltonian $\mathcal{H}_{1}$ in Eq.~(\ref{Hx1}) into the subspace spanned by the above left and right eigenstates as $\mathcal{H}_{\textrm{surf},\alpha\beta}^\textrm{\Rmnum{1}} = \psi_{\alpha,L}^* \mathcal{H}_{1} \psi_{\beta,R}$, we have the effective boundary Hamiltonian in the surface $\textrm{\Rmnum{1}}$ as
\begin{align}\label{Surface1}
	\mathcal{H}_{\textrm{surf,x}}^\textrm{\Rmnum{1}}(k_y, k_z) =   k_y \sigma_z - \left(\eta - \xi\right) \sigma_x,
\end{align}
where we have ignored the terms of order higher than $k_y$, and $\eta$ and $\xi$ are given by
\begin{align}\label{eta}
	\eta =2\Delta_0 \mathcal{N}_1 \mathcal{N}_2 \lambda_{1,R}\lambda_{2,R}/\left(1-\lambda_{1,R}\lambda_{2,R}\right),
\end{align}
\begin{align}\label{xi}
	\xi = \Delta_0 \mathcal{N}_1 \mathcal{N}_2 \lambda_{1,R}\lambda_{2,R} \left(\lambda_{1,R}+\lambda_{2,R}\right)/\left(1-\lambda_{1,R}\lambda_{2,R}\right).
\end{align}

When the system is under open boundary condition along the $y$ direction, and periodic boundary conditions along both the $x$ and $z$ directions, after a partial Fourier transformation along the $k_y$ direction, the Hamiltonian $\mathcal{H}(\mathbf{k})$ in Eq.~(\ref{H1}) becomes
\begin{align}\label{Hy}
	\mathcal{H}_y(k_x, ~k_z) = ~&\sum_{y,k_x,k_z} \Psi_{y,k_x,k_z}^\dagger \left[\left(m_0 - \cos k_x + m_1 \cos k_z \right) s_z \sigma_z    + \left(v_z \sin k_z  + i \gamma\right) s_z  + \sin k_x s_x \sigma_z  \right.  \nonumber \\
	& \left. + \Delta_0 \cos k_x  \sigma_x \right] \Psi_{y,k_x,k_z} +  \sum_{y,k_x,k_z} \left[\Psi_{y,k_x,k_z}^\dagger \left(-\frac{1}{2} s_z \sigma_z   - \frac{i}{2}   s_y \sigma_z  - \frac{\Delta_0}{2} \sigma_x\right)     \Psi_{y+1,k_x,k_z}+ \textrm{H.c.}\right] ,
\end{align}
where $y$ is an integer-valued coordinate taking values from $1$ to $L$. By assuming a small $\Delta_0$, and taking $k_x$ to be close to 0, we rewrite $\mathcal{H}_y$  as $\mathcal{H}_y = \tilde{\mathcal{H}}_{0} + \tilde{\mathcal{H}}_{1}$, with
\begin{align}\label{Hy0}
	\tilde{\mathcal{H}}_{0} = \sum_{y,k_x,k_z} \Psi_{y,k_x,k_z}^\dagger  \tilde{M} \Psi_{y,k_x,k_z}  +  \sum_{y,k_x,k_z} \left(\Psi_{y,k_x,k_z}^\dagger \tilde{T} \Psi_{y+1,k_x,k_z}+ \textrm{H.c.}\right) ,
\end{align}
where $\tilde{M}=\left(m_0 - \cos k_x + m_1 \cos k_z \right) s_z \sigma_z    + \left(v_z \sin k_z  + i \gamma\right) s_z$, and $\tilde{T}=-\frac{1}{2} s_z \sigma_z   - \frac{i}{2}   s_y \sigma_z$, and 
\begin{align}\label{Hy1}
	\tilde{\mathcal{H}}_{1} = ~\sum_{y,k_x,k_z} \Psi_{y,k_x,k_z}^\dagger \left( \sin k_x s_x \sigma_z  + \Delta_0 \cos k_x  \sigma_x \right) \Psi_{y,k_x,k_z} -   \sum_{y,k_x,k_z} \left(\Psi_{y,k_x,k_z}^\dagger \frac{\Delta_0}{2}\sigma_x \Psi_{y+1,k_x,k_z}+ \textrm{H.c.}\right).
\end{align}
where $\tilde{\mathcal{H}}_{1}$ is treated as a perturbation. We first solve the right eigenstates. In order to solve the surface states localized at the boundary $y=L$, we choose a trial solution $\tilde{\psi}_R(y) = \tilde{\lambda}_R^{y} \tilde{\phi}_R$, where $\tilde{\lambda}$ is a parameter determining the localization length with $\abs{\tilde{\lambda}_R}>1$, and $\tilde{\phi}_R$ is a four-component vector. Plugging this trial solution into the Hamiltonian $\tilde{\mathcal{H}}_{0}$ in Eq.~(\ref{Hy0}) for $k_x=0$, we have the following eigenvalue equations:   
\begin{align}\label{sy1}
	\left(\tilde{\lambda}_R^{-1} \tilde{T}^\dagger + \tilde{M} + \tilde{\lambda}_R \tilde{T}\right) \tilde{\phi}_R= E \tilde{\phi}_R, ~~~~~\textrm{in the bulk},
\end{align}
and
\begin{align}\label{sy2}
	\left(\tilde{\lambda}_R^{-1} \tilde{T}^\dagger + \tilde{M}\right) \tilde{\phi}_R= E\tilde{\phi}_R, ~~~~~  \textrm{at the boundary} ~y=L.
\end{align}
By considering the semi-infinite limit along the $y$-axis in a negative direction, and requiring the states have the same eigenenergies in the bulk and at the boundary, we have $\tilde{\lambda}_R \tilde{T} = 0$, which leads to $E=0$, and two corresponding eigenstates $\tilde{\psi}_{1,R}$  and $\tilde{\psi}_{2,R}$.  The eigenstate $\tilde{\psi}_{1,R}$ is written as
\begin{align}\label{eigy1}
	\tilde{\psi}_{1,R} = \tilde{ \mathcal{N}}_1 (\tilde{\lambda}_{1,R} \tilde{\phi}_{1,R},~\tilde{\lambda}_{1,R}^2 \tilde{\phi}_{1,R},~\tilde{\lambda}_{1,R}^3 \tilde{\phi}_{1,R},~\dots),
\end{align}
with
\begin{align}\label{eigy12}
	\tilde{\phi}_{1,R} = (1,~0,~1,~0)^T,~~\textrm{and}~~ \tilde{\lambda}_{1,R} =  1/\left(1 - m_0 - m_1 \cos k_z - v_z \sin k_z  -i \gamma\right),
\end{align}
and the eigenstate $\tilde{\psi}_{2,R}$ is
\begin{align}\label{eigy2}
	\tilde{\psi}_{2,R} = \tilde{ \mathcal{N}}_2 (\tilde{\lambda}_{2,R} \tilde{\phi}_{2,R},~\tilde{\lambda}_{2,R}^2 \tilde{\phi}_{2,R},~\tilde{\lambda}_{2,R}^3 \tilde{\phi}_{2,R},~\dots),
\end{align}
with
\begin{align}\label{eigy22}
	\tilde{\phi}_{2,R} =  (0,~1,~0,~1)^T, ~~\textrm{and}~~ \tilde{\lambda}_{2,R} =  1/\left(1 - m_0 - m_1 \cos k_z + v_z \sin k_z  + i \gamma\right),
\end{align}
here $\lambda_{1,R} = 1/\tilde{\lambda}_{1,R}$ and $\lambda_{2,R} = 1/\tilde{\lambda}_{2,R}$. For the surface states localized at the boundary $y=L$, we require  $\abs{\tilde{\lambda}_{1,R}}>1$ and $\abs{\tilde{\lambda}_{2,R}}>1$, then we have 
\begin{align}\label{eeqqy}
	\left[\left(1 - m_0 - m_1 \cos k_z - v_z \sin k_z\right)^2 + \gamma^2\right]^{1/2}  < 1, ~~\textrm{and}~~~  \left[\left(1 - m_0 - m_1 \cos k_z + v_z \sin k_z\right)^2 + \gamma^2\right]^{1/2}	  < 1.
\end{align}
According to Eq.~(\ref{eeqqy}), as $k_z$ increases from $0$ to $\pi$ (or decreases from $0$ to $-\pi$), the non-Hermitian system $\mathcal{H}_{0}$ first supports two surface states localized at the boundary $y=L$, and then only one surface state localized at the boundary as $\abs{k_z}$ exceeds a critical value (i.e., one of exceptional points at which a phase transition takes place). The critical values of $k_z$ correspond  to ones at which two exceptional rings closest to $k_z=0$ locate for the case of   small $\gamma$. Because the domain-wall states, as discussed below, only appear if $\mathcal{H}_{0}$ supports two surface states, the hinge Fermi-arc states exist only for a finite regime of $k_z$.

For the left eigenstates under the open boundary condition along the $y$ direction, we assume a trial solution $\tilde{\psi}_L(y) = \tilde{\lambda}_L^{y} \tilde{\phi}_L$. Considering the same procedure for deriving the right eigenstates, we obtain the left eigenstates $\tilde{\psi}_{1,L}$  and $\tilde{\psi}_{2,L}$ as  
\begin{align}\label{eigy1L}
	\tilde{\psi}_{1,L} = \tilde{ \mathcal{N}}^*_1 (\tilde{\lambda}_{1,L} \tilde{\phi}_{1,L},~\tilde{\lambda}_{1,L}^2 \tilde{\phi}_{1,L},~\tilde{\lambda}_{1,L}^3 \tilde{\phi}_{1,L},~\dots),
\end{align}
with
\begin{align}\label{eigy12L}
	\tilde{\phi}_{1,L} = (1,~0,~1,~0)^T,~~\textrm{and}~~ \tilde{\lambda}_{1,L} =  1/\left(1 - m_0 - m_1 \cos k_z - v_z \sin k_z  +i \gamma\right),
\end{align}
and 
\begin{align}\label{eigy2L}
	\tilde{\psi}_{2,L} = \tilde{ \mathcal{N}}^*_2 (\tilde{\lambda}_{2,L} \tilde{\phi}_{2,L},~\tilde{\lambda}_{2,L}^2 \tilde{\phi}_{2,L},~\tilde{\lambda}_{2,L}^3 \tilde{\phi}_{2,L},~\dots),
\end{align}
with
\begin{align}\label{eigy22L}
	\tilde{\phi}_{2,L} =  (0,~1,~0,~1)^T, ~~\textrm{and}~~ \tilde{\lambda}_{2,L} =  1/\left(1 - m_0 - m_1 \cos k_z + v_z \sin k_z  - i \gamma\right),
\end{align}

For the $k_z$ region where the system supports two surface states, projecting the Hamiltonian $\tilde{\mathcal{H}}_{1}$ in Eq.~(\ref{Hy1}) into  the subspace spanned by the above right and left eigenstates as $\mathcal{H}_{\textrm{surf},\alpha\beta}^\textrm{\Rmnum{4}} = \tilde{\psi}_{\alpha,L}^* \tilde{\mathcal{H}}_{1} \tilde{\psi}_{\beta,R}$, we have the effective boundary Hamiltonian in the surface $\textrm{\Rmnum{4}}$
\begin{align}\label{Surface4}
	\mathcal{H}_{\textrm{surf,y}}^\textrm{\Rmnum{4}}(k_x, k_z) =   k_x \sigma_z + \left(\eta - \xi \right) \sigma_x,
\end{align}
where we have ignored the terms of order higher than $k_x$.

By using the same procedure above, we have the effective  Hamiltonian for boundary states localized at surfaces $\textrm{\Rmnum{2}}$ and $\textrm{\Rmnum{3}}$ as
\begin{align}\label{Surfacey2}
	\mathcal{H}_{\textrm{surf,y}}^\textrm{\Rmnum{2}}(k_x, k_z) =   -k_x \sigma_z + \left(\eta - \xi \right) \sigma_x,
\end{align}
\begin{align}\label{Surface3}
	\mathcal{H}_{\textrm{surf,x}}^\textrm{\Rmnum{3}}(k_y, k_z) =   -k_y \sigma_z - \left(\eta - \xi \right) \sigma_x.
\end{align}

According to the surface Hamiltonians in Eqs.~(\ref{Surface1}) and (\ref{Surface4}-\ref{Surface3}), as well as the condition in Eq.~(\ref{eeqq}), the surface states, in the gapped regime and a finite $k_z$ region, show the same kinetic energy coefficients, but the mass terms on two neighboring surfaces always have opposite signs. Therefore, the mass domain walls appear at the intersection of two neighboring surfaces, and these two surfaces can share a common zero-energy boundary state (analogous to the Jackiw-Rebbi zero modes \cite{PhysRevD.13.3398SM}) in spite of complex-valued $\mu$, which corresponds to the hinge Fermi-arc states at each $k_z$.  Moreover, these hinge Fermi-arc states exist only in a finite $k_z$ region limited by the condition in Eq.~(\ref{eeqq}). This explains why the Hamiltonian $\mathcal{H}(\mathbf{k})$ shows both first-order and higher-order topological features for small $\gamma$.

\begin{figure}[!t]
	\centering
	\includegraphics[width=17.8cm]{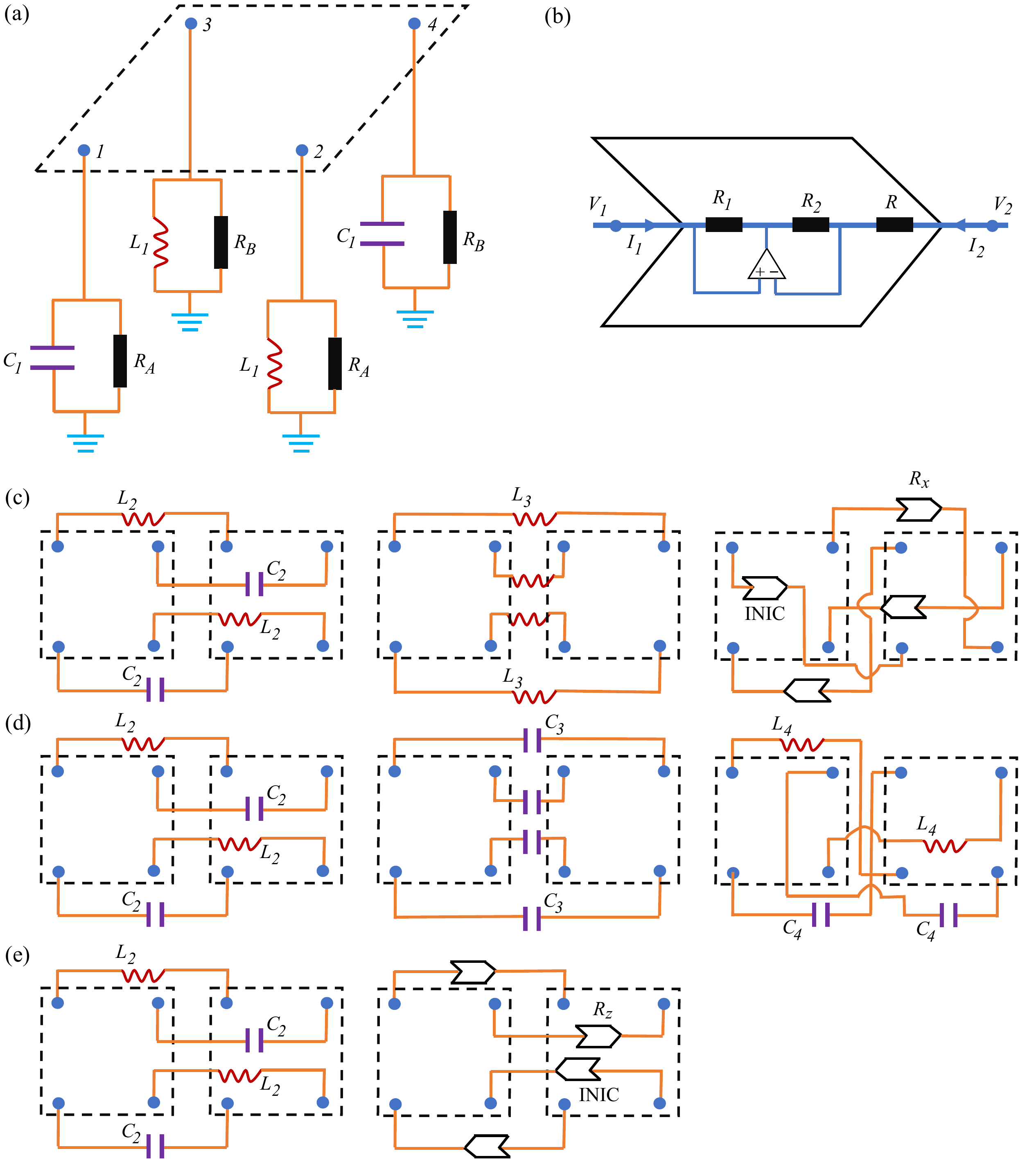}
	\caption{ (a) Schematic diagram of a unit cell of a cubic lattice realized by electric circuits. The unit-cell electric circuits consist of four nodes, and each node is connected to grounded electric elements for simulating the diagonal entries in the Hamiltonian $\mathcal{H}_1(x, ~y,~z)$. The on-site gain and loss are realized by resistive elements $R_A$ and $R_B$. $C_1$ and $L_1$ denote capacitances and inductances. (b) Negative impedance converter  with current inversion (INIC) used for the hopping with imaginary amplitudes. (c-e) Diagrams of the electric circuits for simulating Hamiltonians $\mathcal{H}_2(x, ~y,~z)$, $\mathcal{H}_3(x, ~y,~z)$, and $\mathcal{H}_4(x, ~y,~z)$, respectively. Here, $C_2$, $C_3$ and $C_4$ are capacitances, $L_2$, $L_3$ and $L_4$ denote inductances, $R_x$ and $R_z$ represent resistances of INICs. }\label{circSM}
\end{figure}

\section{Possible Experimental Realizations Using topoelectric circuits }

Recently, non-Hermitian first-order  and higher-order topological insulators \cite{Helbig2020SM,arXiv:2104.11260SM}, 3D Hermitian higher-order topological insulators  \cite{Liu2020SM} and topological semimetals \cite{Lee2020SM} have been experimentally observed in topoelectric circuits. These indicate that  electric circuits are excellent platforms to realize complicated and exotic topological structures. In this section, we propose to realize the lattice model in Eq.~(1) in the main text using topoelectric circuits. Without loss of generality, we set $m_1=-1$, $m_0>0$, $v_z>0$ and $\gamma>0$.

The real-space Hamiltonian $\mathcal{H} (x, ~y,~z)$ for Eq.~(1) in the main text reads
$\mathcal{H} (x, ~y,~z) = \mathcal{H}_1 (x, ~y,~z)+\mathcal{H}_2 (x, ~y,~z)+\mathcal{H}_3 (x, ~y,~z)+\mathcal{H}_4(x, ~y,~z)$, where
\begin{align}\label{real_1}
	\mathcal{H}_1(x, ~y,~z) =\sum_{x, y, z} \Psi_{x, y, z}^\dagger \left( m_0  s_z \sigma_z    + i \gamma s_z   \right)\Psi_{x, y, z},
\end{align}
\begin{align}\label{real_2}
	\mathcal{H}_2(x, ~y,~z) =  &\sum_{x, y, z} \left[\Psi_{x, y, z}^\dagger \left(-\frac{1}{2} s_z \sigma_z   - \frac{i}{2}   s_x \sigma_z  + \frac{\Delta_0}{2} \sigma_x\right)     \Psi_{x+1, y, z}+ \textrm{H.c.}\right],
\end{align}
\begin{align}\label{real_3}
	\mathcal{H}_3(x, ~y,~z) =  \sum_{x, y, z} \left[\Psi_{x, y, z}^\dagger \left(-\frac{1}{2} s_z \sigma_z   - \frac{i}{2}   s_y \sigma_z  - \frac{\Delta_0}{2} \sigma_x\right)     \Psi_{x, y+1,  z}+ \textrm{H.c.}\right],
\end{align}
and
\begin{align}\label{real_4}
	\mathcal{H}_4(x, ~y,~z) = ~&\sum_{x,y,z} \left[ \Psi_{x,y,z}^\dagger \left(-\frac{1}{2} s_z \sigma_z- i\frac{v_z}{2} s_z \right)    \Psi_{x,y,z+1} + \textrm{H.c.}\right].
\end{align}

We now consider a 3D electric-circuit network forming a cubic lattice. The electric-circuit network consists of various nodes labeled by $a$. According to Kirchhoff's law, the current $I_a$ entering the circuit at a node $a$ equals the sum of the currents $I_{ab}$ leaving it to other nodes or ground
\begin{align}\label{kir_law}
	I_a = \sum_{b} I_{ab} = \sum_{b} X_{ab} (V_a-V_b) + X_a V_a,
\end{align}
where $X_{ab} = 1/Z_{ab}$ is the admittance between nodes $a$ and $b$ ($Z_{ab}$ is the corresponding impedance), $X_a$ is the admittance between node $a$ and the ground, and $V_a$ is voltage at node $a$. Using Eq.~(\ref{kir_law}), the external input current $I_a$ and the node voltage $V_a$ can be rewritten into the following matrix equation
\begin{align}\label{lapcian}
	\mathbf{I}(\omega) = \mathbf{J}(\omega) \mathbf{V}(\omega),
\end{align}
where $\mathbf{I}= (I_1,~ I_2, \cdots, ~I_N)$, $\mathbf{V}= (V_1,~ V_2, \cdots, ~V_N)$, and $N$ is the physical dimension. Here the $N\times N$ matrix $\mathbf{J}(\omega)$ is the circuit Laplacian, which can be used to simulate the system Hamiltonian $\mathcal{H}(\mathbf{k})$, having the form \cite{Lee_2018SM, Tao_2020SM,PhysRevResearch.3.023056SM}
\begin{align}\label{lapcian2}
	\mathbf{J}(\omega) = i \omega \mathcal{L}(\omega)  = i \omega \mathbf{C} + \frac{1}{i \omega \mathbf{L}} + \frac{1}{\mathbf{R}},
\end{align}
where $\mathbf{C} $, $\mathbf{L}$ and $\mathbf{R}$  are the capacitance, inductance and resistance matrices, respectively.

To simulate the Hamiltonian $\mathcal{H} (x, ~y,~z)$, we require $\mathcal{L} =\mathcal{H}$.  Figure \ref{circSM}(a) plots the unit-cell circuit for the cubic lattice consisting of four nodes. Each node is connected to grounded electric elements for simulating the diagonal entries in the Hamiltonian $\mathcal{H}_1(x, ~y,~z)$ in Eq.~(\ref{real_1}). The on-site gain and loss are realized by the resistive elements $R_A$ and $R_B$. The electric circuits for simulating Hamiltonians $\mathcal{H}_2(x, ~y,~z)$, $\mathcal{H}_3(x, ~y,~z)$, and $\mathcal{H}_4(x, ~y,~z)$ are shown in Fig.~\ref{circSM}(c-e). The inductors and capacitors between two neighboring nodes contribute hopping terms with positive and negative amplitudes \cite{Tao_2020SM}, respectively. For the hopping with imaginary amplitude, we use a negative impedance converter with current inversions (INICs) \cite{Tao_2020SM}, as shown in Fig.~\ref{circSM}(b). When the current flows towards the INICs (the large arrow), the resistance is negative, and it is positive when the direction is opposite.

As indicated in the electric circuits in Fig.~\ref{circSM}, the Laplacian that simulates the Hamiltonian $\mathcal{H}(\mathbf{k})$ in Eq.~(1) in the main text reads
\begin{align}\label{simulation}
	\mathcal{L}(\omega) = ~&\left(M_E - J_E \cos k_x  - J_E \cos k_y  - J_E \cos k_z \right) s_z \sigma_z    + \left(v_E \sin k_z  + i \gamma_E \right) s_z +  t_{E1} \sin k_x   s_x \sigma_z  + t_{E2} \sin k_y s_y \sigma_z  \nonumber \\ 
	& + \Delta_E  \left(\cos k_x - \cos k_y \right) \sigma_x + i \lambda_E \mathcal{I},
\end{align}
where
\begin{align}\label{s221}
	M_E = C_1 = 1/(\omega^2 L_1), ~~\gamma_E=1/(2 \omega R_B) - 1/(2 \omega R_A), 
\end{align}
\begin{align}\label{s222}
	\lambda_E=-1/(2 \omega R_B) - 1/(2 \omega R_A), ~~J_E=C_2/2=1/(2\omega^2 L_2), 
\end{align}
\begin{align}\label{s223}
	v_{E} = 1/(2 \omega R_z),  ~~t_{E1} = 1/(2 \omega R_x), ~~t_{E2} = C_4/2=1/(2\omega^2 L_4), ~~\Delta_E=C_3/2=1/(2\omega^2 L_3),
\end{align}
and $\mathcal{I}$ is identity matrix. Note that the last non-Hermitian term does not change the topological features of the system. This electric circuit can be utilized to investigate the non-Hermitian higher-order Weyl-exceptional-ring semimetals studied in this work.

\end{document}